\documentclass{Interspeech2024}
\usepackage{subfigure}

\interspeechcameraready

\title{Expressive paragraph text-to-speech synthesis with multi-step variational autoencoder}

\name{Xuyuan} {Li$^{\spadesuit \dagger \star}$}
\name{Zengqiang} {Shang$^{\diamondsuit \spadesuit \star}$}
\name{Peiyang} {Shi$^{\spadesuit}$}
\name{Hua} {Hua$^{\spadesuit \dagger}$}
\name{Ta} {Li$^{\spadesuit \dagger}$}
\name{Pengyuan} {Zhang$^{\diamondsuit \spadesuit \dagger}$}

\address{$^{\spadesuit}$  Laboratory of Speech and Intelligent Information Processing, Institute of Acoustics, CAS, China \\
      $^{\dagger}$ University of Chinese Academy of Sciences, China}

\email{[lixuyuan, shangzengqiang, shipeiyang, huahua, lita, zhangpengyuan]@hccl.ioa.ac.cn}


\begin{document}

\maketitle

\begin{abstract}
    
Neural networks have been able to generate high-quality single-sentence speech. However, it remains a challenge concerning audio-book speech synthesis due to the intra-paragraph correlation of semantic and acoustic features as well as variable styles. In this paper, we propose a highly expressive paragraph speech synthesis system with a multi-step variational autoencoder, called EP-MSTTS. EP-MSTTS is the first VITS-based paragraph speech synthesis model and models the variable style of paragraph speech at five levels: frame, phoneme, word, sentence, and paragraph. We also propose a series of improvements to enhance the performance of this hierarchical model. In addition, we directly train EP-MSTTS on speech sliced by paragraph rather than sentence. Experiment results on the single-speaker French audiobook corpus released at Blizzard Challenge 2023 show EP-MSTTS obtains better performance than baseline models.
 
\end{abstract}

\renewcommand{\thefootnote}{\fnsymbol{footnote}}
\footnotetext{$^{\star}$ Equal contribution.}
\renewcommand{\thefootnote}{\fnsymbol{footnote}}
\footnotetext{$^{\diamondsuit}$ Corresponding author.}
\setcounter{footnote}{0}
\renewcommand{\thefootnote}{\arabic{footnote}}

\section{Introduction}

Thanks to the development of neural networks, current text-to-speech (TTS) models \cite{ren2020fastspeech, kong2020hifi, kim2021conditional, shang2021incorporating} have achieved near-human synthesis results on customized datasets recording independent sentences. However, it remains a challenge for synthesizing audiobooks. Different from the customized datasets, audiobooks have two special characteristics: 1) The speech of audiobooks is usually given in the form of paragraphs because the semantic and acoustic features (unrelated to semantics such as acoustic environment, hoarse voice, and nasality) of different sentences in the same paragraph are strongly correlated. 2) There are dramatic shifts in style within a single paragraph, as the reader needs to switch between styles to play the role of narrator and different characters. Recent research \cite{rodero2023synthetic} points out that TTS methods still have obvious drawbacks when used to synthesize long stories in audiobooks compared with humans.

There are many works have been devoted to capturing semantic dependence. \cite{lei2022towards,xu2021improving,li2022enhancing} used the embedding of the context sentences
extracted by pre-trained language model \cite{devlin2018bert,yang2019xlnet}
to predict the global style of the current sentence. In addition to the global style, \cite{chen2022unsupervised} additionally models local styles based on context semantics. StyleSpeech \cite{chen2023stylespeech} followed this two-stage style modeling and used unpaired text and audio data to pre-train the context encoder and reference encoder to improve the generalization. However, these works are trained at the sentence level and ignore modeling the acoustic features which are independent of context semantics. These acoustic features are generally consistent across paragraph-level speech but change throughout the dataset due to the change of record location and the physiological state of speakers. Thus, given different sentences in a paragraph, these sentence-level models could generate speech with very different acoustic features. To generate speech with consistent acoustic features \cite{xiao2023contextspeech,lei2023context,xin2023improving} considered previously synthesized speech when synthesizing the current sentence. Another solution is training the model directly on the whole paragraph speech. ParaTTS \cite{xue2022paratts} first performed this paragraph-level training on a modified Tacotron2 \cite{shen2018natural}. \cite{makarov22_interspeech} applied it on Fastspeech2 \cite{ren2020fastspeech} and found paragraph-level training has a huge advantage over single-sentence training in the following two cases: 1) Paragraphs containing many very short sentences. 2) Paragraphs with dramatic shifts in style. Another recent work \cite{zhang2023audiobook} found that increasing the duration limit of this long-form training can further enhance the performance of paragraph speech synthesis model.

However, these models are based on FastSpeech or Tacotron, which performs not as well as VITS \cite{kim2021conditional}. In addition, they model the style of paragraph speech at no more than two levels of hierarchy, leading to average rhythms. In this paper, we work on adapting the VITS to generate paragraph speech with highly expressive. To achieve this, we expand the single-layer variational autoencoder (VAE) of the original VITS model into five layers, inspired by HierTTS \cite{shang2023hiertts}. We describe how it captures the variable style of paragraph speech at five levels: frame, phoneme, word, sentence, and paragraph in Section \ref{sec:format}. After that, in Section \ref{sec:ts}, we propose a novel training strategy to alleviate the posterior collapse problem of deep VAE. Finally, we apply paragraph-level training to our proposed model to generate paragraph speech with smooth transitions in style and acoustic features. The subjective and objective evaluation results on a French audiobook dataset show that our proposed model outperforms all baseline models. The results of ablation studies also demonstrate the effectiveness of our proposed module and training strategy. Audio can be found at https://lixuyuan102.github.io/EP-MSTTS/.

\begin{figure*}[htbp]
\centering
\includegraphics[width=0.9\textwidth]{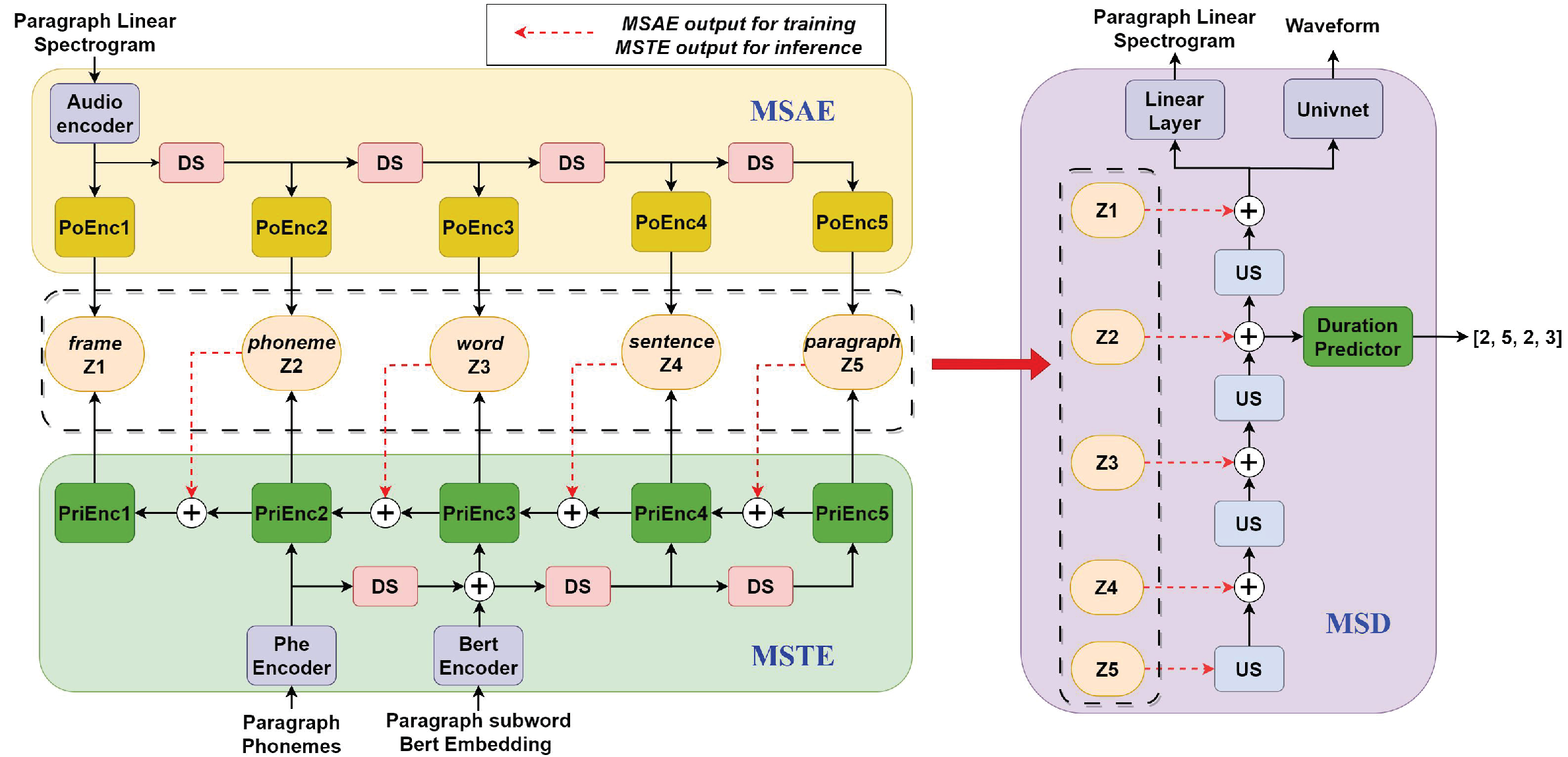}
\caption{EP-MSTTS architecture including MSAE, MSTE, and MSD. The DS and US denote "Downsampling" and ”Upsampling" operations. The PoEnc and PriEnc denote "Posterior Encoder" and "Prior Encoder". "$\oplus$" represents concatenating two tensors.}
\label{fig:1}
\end{figure*}

\section{Model Structure}
\label{sec:format}

For convenience, we refer to our proposed model as EP-MSTTS. The overall architecture of EP-MSTTS can be divided into three modules: Multi-step Audio Encoder (MSAE), Multi-step Text Encoder (MSTE), and Multi-step Decoder (MSD), as illustrated in Fig. \ref{fig:1}. All modules have five layers including $L_1$, $L_2$, $L_3$, $L_4$, and $L_5$, which represent the frame, phoneme, word, sentence, and paragraph layer, from fine to coarse.

\subsection{Multi-step Audio Encoder}
\label{ssec:msae}
Within a paragraph of audiobooks, different characters usually have different styles of sentences. The speaker usually expresses these styles by regularly changing the pitch and volume of words. To capture the variable style and express it well, the MSAE breaks down a paragraph speech at five grammatical levels. Specifically, as shown in the upper left (yellow) part of Fig. \ref{fig:1}, we downsample the hidden state of paragraph linear-spectrogram step by step according to the pre-given alignment. The acoustic hidden state $\bm{h_{i}^{au}}$ at each step is encoded into a posterior latent variable by posterior encoder $\mathrm{G_{post}}$:
\begin{equation}
    \bm{u_i}, \bm{\sigma_i} = \mathrm{G_{post}}(\bm{h_{i}^{au}})\label{eq:1}
\end{equation}
where $\bm{u_i}$ and $\bm{\sigma_i}$ represent the means and variances of the posterior latent variables. Since the information in the initial step is much larger than in the subsequent steps. We only employ the Posterior Encoder mentioned in VITS \cite{kim2021conditional} as the audio encoder of the frame layer. For the other levels, we follow the HierTTS \cite{shang2023hiertts}, which uses a bidirectional GRU with attention-pooling as the downsample operation to extract coarser-grained information from the hidden state of the upper layer. All the posterior encoders are composed of a convolution layer whose output is activated by Softplus. We emphasize that the MSAE is different from the Hierarchical Audio Encoder (HAE) in HierTTS, which uses text information as an additional input. To avoid posterior latent variables learning useless information directly from the text, our MSAE only learns information from audio.

\vspace{-0.1cm}
\subsection{Multi-step Text Encoder}
\label{ssec:mste}

Coarse-to-fine iterative generation can produce better results than single-step generation\cite{gregor2015draw}. Based on this principle, we followed the structure of the Hierarchical Audio Decoder from \cite{shang2023hiertts} and modified its layer depth to fit the grammatical hierarchy of French paragraphs, as shown in the lower left (green) part of Fig. \ref{fig:1}. Starting from the paragraph level, the MSTE encodes semantic information to the prior latent variables. The specific operation of each step is written as follows:
\begin{equation}
    \bm{s_i}, \bm{u_i}, \bm{\sigma_i} = \mathrm{G_{pri}}(\mathrm{US}(\bm{s_{i-1}} + \bm{z_{i-1})},\bm{h_{i}^{txt}})\label{eq:2}
\end{equation}
where prior encoder $\mathrm{G_{pri}}$ predicts the means $\bm{u_i}$ and variances $\bm{\sigma_i}$ of prior latent variables from semantic information $\bm{h_{i}^{txt}}$, posterior latent variables $\bm{z_{i-1}}$ from the previous step, and the hidden state of prior predictor $\bm{s_{i-1}}$ from the previous step. $\mathrm{US}$ means Upsampling operation, which will be described in Sec. \ref{ssec:msd}. We use a lookup table to obtain the hidden state of paragraph phonemes, two convolution layers with a kernel size of 3, and a Downsampling module to extract the hidden state of words from the Bert subword embedding. The Downsampling operation is implemented in the same way as in MSAE. Because semantic information cannot be split to the frame level, we do not set semantic information input for the frame layer. All prior encoders consist of $N$ feed-forward transformer blocks \cite{ren2019fastspeech} and a linear layer. Following the order from the frame layer to the paragraph layer, we set $N$ to [4,4,3,3,2] respectively.

\subsection{Multi-step Decoder}
\label{ssec:msd}
We hypothesize that generating speech waveforms using only $\bm{z}$ from the frame layer leads to redundancy of information between layers and accumulates inference errors across layers, since MSTE inference $\bm{z}$ in a recursive manner. To address this, we upsample the hidden states $\bm{h}$ of the upper step and concatenate it with $\bm{z}$ at each step, as illustrated in the right (purple) part of Fig. \ref{fig:1}. The Upsampling operation is implemented by the Length Regulator in \cite{ren2019fastspeech}. The hidden states at each step are given by equation (\ref{eq:3}).
\begin{equation}
    \bm{h_i} = \mathrm{US}(\bm{h_{i-1})} + \bm{z_i}\label{eq:3}
\end{equation}
We employ the Univnet \cite{jang2021univnet} to generate speech waveforms from the hidden states in the last step. Since the 
$\bm{z}$ of each layer can contribute directly to the reconstruction of the speech waveform. Encoders tend to encode independent information into different layers. For vocoder, this decoding method also mitigates the mismatch between training and inference caused by multi-layer cumulative errors. In addition, we employ a linear layer as the parallel structure of the vocoder to predict the linear-spectrogram of the paragraph speech. The specific effect of this parallel structure is described in section \ref{sec:ts}.

\section{Training Strategy}
\label{sec:ts}

We employ KL loss to make the output of MSTE consistent with that of MSAE. The KL loss $\mathcal{L}_{kl}$ of all 
 grammatical layers can be expressed as:
\begin{equation}
\mathcal{L}_{kl} = \alpha_1 \mathcal{L}_1 + \alpha_2 \mathcal{L}_2 + \alpha_3 \mathcal{L}_3 + \alpha_4 \mathcal{L}_4 + \alpha_5 \mathcal{L}_5 \label{eq:4}
\end{equation}
We set the $\alpha_1$, $\alpha_2$, $\alpha_3$, $\alpha_4$, and $\alpha_5$ as a decreasing sequence: [1, 0.25, 0.07, 0.01, 0.005] because the information from the finer-grained layer contributes more to speech reconstruction. For MSD, we apply the adversarial training on vocoder with Multi-Period Discriminator \cite{kong2020hifi} and Multi-Resolution Discriminator \cite{jang2021univnet}. The adversarial loss is denoted by $\mathcal{L}_g$ . We apply the L1 loss of mel-spectrogram $\mathcal{L}_{m}$ and multi-resolution STFT loss \cite{yamamoto2020parallel} $\mathcal{L}_{s1}$ on the output of the vocoder. In addition, another multi-resolution STFT loss $\mathcal{L}_{s2}$ is used to constrain the output of the linear layer parallel to the vocoder. Finally, we use the L2 loss $\mathcal{L}_{d}$ to train the duration predictor. 

We observed that when directly training the total model, Univnet tends to reconstruct waveform only from the $\bm{z}$ at low levels due to its overpowering reconstruction capability. This phenomenon, known as the posterior collapse, underscores the necessity for a less potent vocoder to force the encoder to supply more useful information. However, restructuring the waveform is difficult, and we cannot risk compromising the quality of the speech reconstruction by choosing a simple vocoder structure. Inspired by MSMC-TTS \cite{guo2023msmc}, we designed another task: predicting the linear-spectrogram with a simple linear layer parallel to the vocoder. We also use the weight annealing \cite{bowman-etal-2016-generating} on the weight of $\mathcal{L}_{kl}$ to ensure that our model has enough time to extract information from audio without text constraints. Specifically, our training strategy is divided into the following three stages:
\begin{enumerate}
    \item First, only the MSAE and MSD are trained with the goal of reconstructing the linear spectrogram features. The weight $\lambda_{kl}$ of $\mathcal{L}_{kl}$ is set to 0.00001. As a result, the total loss can be given as:
\begin{equation}
    \mathcal{L}_{tot} = 2.5\mathcal{L}_{s2} + 5\mathcal{L}_{d} + \lambda_{kl} \mathcal{L}_{kl}  \label{eq:5} 
\end{equation}
    \item Then the $\lambda_{kl}$ starts to increase slowly, which means the latent variable of each layer begins to adjust the style information it contains with the constraints of the MSTE. We change the $\lambda_{kl}$ as following equation:
\begin{equation}
    \lambda_{kl} = 1e^{-5}(Step_g - Step_w)
\end{equation}
    $Step_g$ is the global training iterations. $Step_w$ is the total number of iterations for the first stage.
    \item Finally, the network gives up reconstructing the linear spectrogram features and instead reconstructs the speech waveform. Meanwhile, $\lambda_{kl}$ keeps increasing until 1. As a result, the total loss is given as:
\begin{equation}
    \mathcal{L}_{tot} = \mathcal{L}_g + 1.5 \mathcal{L}_{s1} + 2.5 \mathcal{L}_{m} + \lambda_{kl} \mathcal{L}_{kl} + \mathcal{L}_{d}  \label{eq:6} 
\end{equation}
\end{enumerate}
The first and second stages were set to 10,000 and 30,000 iterations, respectively.

\section{Experiments}
\label{sec:pagestyle}

\subsection{Dataset}
\label{ssec:dataset}

\begin{figure}[ht]
\centering
\includegraphics[width=0.47\textwidth]{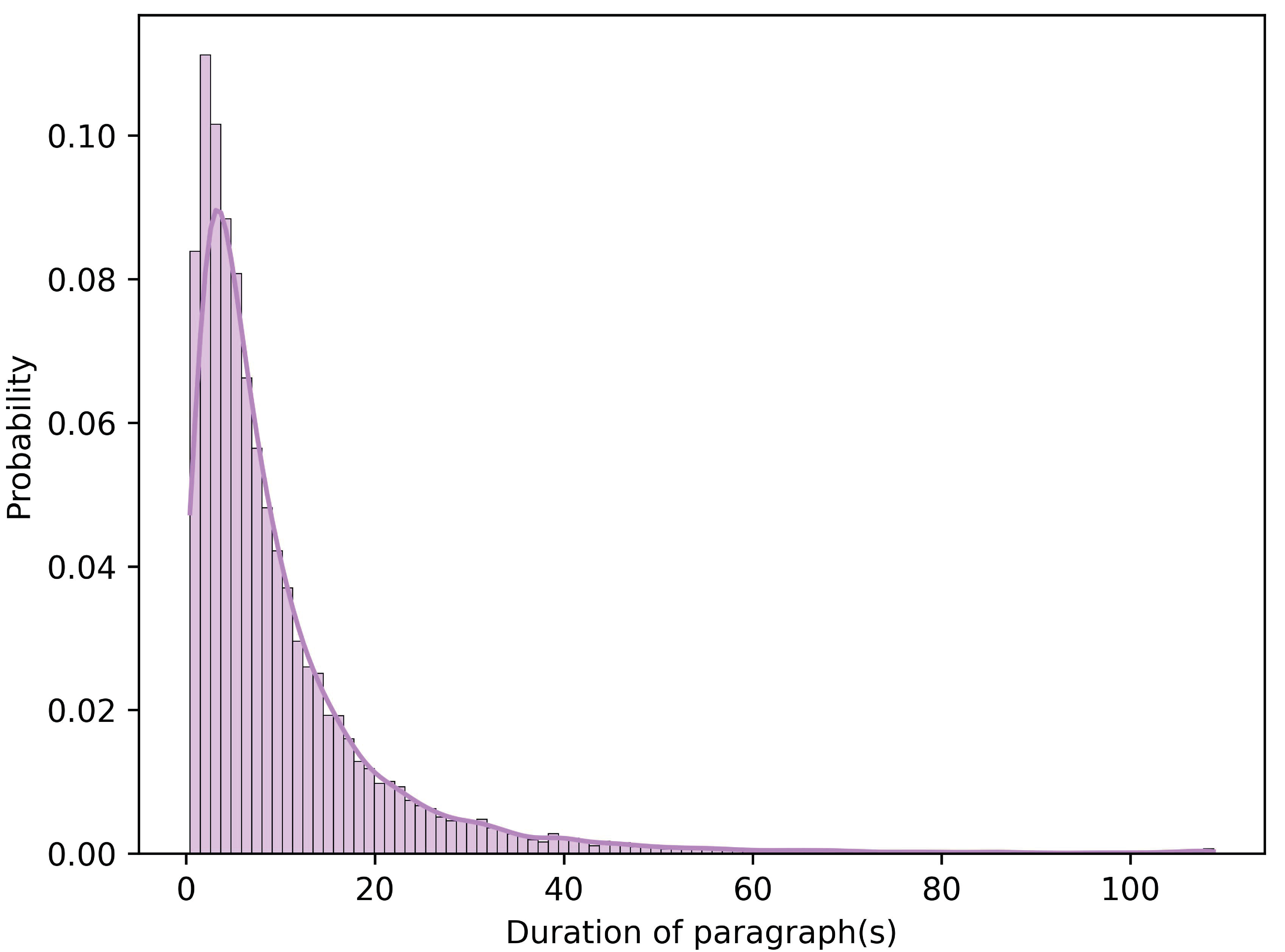}
\caption{Probability distribution of paragraph duration length.}
\label{fig:2}
\end{figure}

In this work, we used a portion of the French audiobook dataset \cite{baillygerard20237560290}. This part consists of 289 chapters of 5 audiobooks from Librivox read by a female native French speaker. After slicing, we compiled 18,155 paragraphs \footnote{We cut 26 paragraphs that exceeded 218 seconds into smaller segments that did not exceed 218 seconds.} totaling 51.6 hours. The length distribution of these paragraph speech is shown in Fig \ref{fig:2}. All the audios are encoded in the PCM format at a sampling rate of 22.05 kHz with transcribed text and partial aligned phoneme sequences. We used a transformer-based sequence-to-sequence model to generate the remaining phonemes and used the MFA \cite{mcauliffe2017montreal} to align these phonemes and audios. In our experiments, 150 paragraphs were selected for testing, 150 paragraphs for validation, and the remaining paragraphs for training. We extracted the 512-dimensional linear-spectrograms with a hop length of 300 points and a window length of 800 points. A pre-trained Bert released on Huggingface\footnote{https://github.com/huggingface/transformers} was used to extract the subword embedding for paragraph text. 

\subsection{Experiment settings}
\label{ssec:exp_set}

Four models were implemented as baseline models. 1) VITS-U: To demonstrate that our VITS-based model is more suitable than the original VITS for synthesizing paragraph speech in audiobooks. We cloned the official implementation\footnote{https://github.com/jaywalnut310/vits} of VITS as one of the baseline models. To eliminate the bias introduced by the vocoder, we replaced the vocoder structure in VITS with Univnet\footnote{https://github.com/rishikksh20/UnivNet-pytorch}. 2) VITS-U-S: We split the audios in the dataset by sentence and trained the VITS-U on them, to compare the performance of sentence-level training with paragraph-level training on VITS. 3) VITS-U-B: We add a style encoder for VITS to represent models that utilize semantic information to predict the style of paragraph speech at a limited number of hierarchical levels. The style encoder first upsample the Bert subword embedding to the word level, as described in Sec \ref{ssec:mste}, then downsample it to the phoneme level, and finally converts it to a style embedding utilizing the same structure as the text encoder in VITS. VITS-U-B concatenates this style embedding and text embedding together to predict the prior latent variables. 4) HierTTS: To demonstrate that our proposed improvements enhance the hierarchical modeling. We cloned the official implementation\footnote{https://github.com/shang0712/HierTTS} of HierTTS and modified it to the same five-layer structure as EP-MSTTS. All models except VITS-U-S are trained at the paragraph level using a dynamic batch with a total duration of no more than 218 seconds.

\begin{figure}[htbp]
\centering
\includegraphics[width = 0.8\linewidth]{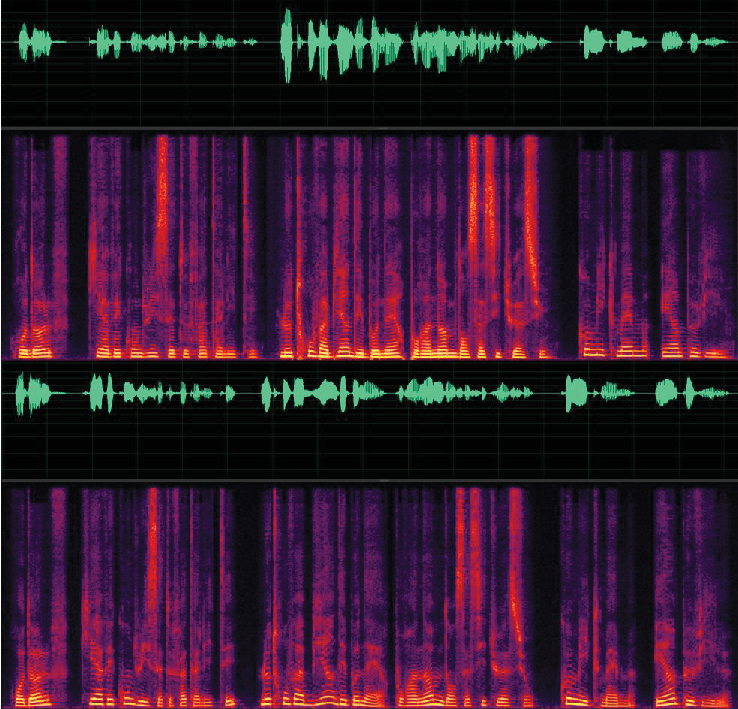}
\caption{Speech of three continuous sentences generated by VITS-U-S (top) and VITS-U (bottom).}
\label{fig:3}
\end{figure}
\vspace{-0.2cm}

\subsection{Evaluation}
\label{ssec:eval}

\subsubsection{Training Evaluation}
We visualized the speech of three continuous sentences generated by VITS-U-S and VITS-U. The speech of VITS-U-S is generated sentence by sentence and is concatenated together, while the speech of VITS-U is generated in a single inference. As shown in Fig. \ref{fig:3}, besides its inability to predict the pauses between sentences, VITS-U-S generates speech with discontinuities in loudness (the second sentence) and high frequency (the third sentence). On the contrary, the acoustic features of the speech synthesized by VITS-U are consistent. In addition, we further computed the mel cepstral distortion (MCD) and Log-F0 root mean square error (log-F0 RMSE) of generated speech with the scripts in ESPnet\footnote{https://github.com/espnet/espnet/tree/master/egs2/TEMPLATE/tts1}. As shown in Tab. \ref{tab1}, compared to VITS-U-S, VITS-U decreased by 0.034 on MCD and remained unchanged on log-F0 RMSE. These results illustrate that paragraph-level training leads to better paragraph audiobook synthesis on VITS, which is consistent with the findings of previous experiments on FastSpeech-based and Tacotron-based models.

\begin{table}[htbp]
\small
\caption{Subjective and objective comparison on test set}
\label{tab1}
\resizebox{0.99\hsize}{!}{$
\centering
\begin{tabular}{c|c|cc}
\toprule [2pt]
Metric & MOS (±95\%CI) $\uparrow$ &  MCD $\downarrow$ & log-F0 RMSE $\downarrow$ \\
\midrule 
GT        & 4.21±0.07   & N/A  & N/A \\
VITS-U-S & \textbf{-} & 5.692 &0.266 \\
VITS-U   & 3.35±0.09 & 5.658 &0.266 \\
VITS-U-B & 3.52±0.07   & 5.574 &0.246 \\
HierTTS    & 3.80±0.08 & 5.458 &0.250 \\
EP-MSTTS  & \textbf{3.98±0.07}   & \textbf{4.876} &\textbf{0.215}\\
\bottomrule [2pt]     
\end{tabular}
$}
\end{table}

\subsubsection{Model Evaluation}
\label{ssec:sub_eval}
We perform subjective and objective evaluations on models that are trained at the paragraph level. 25 native French speakers were invited to score as the role of the 5-scale mean opinion score (MOS) in terms of sound quality, naturalness, and expressiveness. Each listener was randomly assigned 20 paragraphs from the test set. As Tab. \ref{tab1} shown, VITS-U-B outperforms VITS-U on both subjective and objective metrics, which demonstrates the importance of contextual semantic information for capturing the style of paragraph speech. In addition, HierTTS, which models the style with five grammatical levels, outperforms the single-level VITS-U-B on MOS and MCD, but its Log-F0 RMSE rises by a relative 0.004. This is reasonable because, as described in its original paper \cite{shang2023hiertts}, the F0 of speech generated by HierTTS fluctuates over a wide range. We believe this is mainly due to the recursive five-layer decoder accumulating inference errors at each layer. Finally, our proposed model obtains the highest MOS and has significant reductions over HierTTS on MCD/log-F0 with 0.582/0.035. These results illustrate that our proposed modifications significantly improve the performance of the VITS-based s hierarchical model on audiobook paragraph speech synthesis.

\begin{table}[htbp]
\caption{CMOS results on ablation studies.}
\centering
\label{tab2}
\begin{tabular}{l|c}
\toprule [2pt]
\multicolumn{1}{c|}{Models} & CMOS\\ 
\midrule 
EP-MSTTS  & 0\\
+Text in MSAE  & -0.12  \\
-MSD   & -0.26\\
-Training Strategy& -0.17  \\
\bottomrule [2pt]     
\end{tabular}
\vspace{-0.1cm}
\end{table}

\subsubsection{Ablation Study}
\label{ssec:ab_exp}
EP-MSTTS differs from previous Hierarchical VAE models mainly in MSAE, MSD, and the training strategy. We conducted the comparative MOS (CMOS) test (on a 7-point scale) to demonstrate the effectiveness of these proposed modifications. As shown in Tab. \ref{tab2}: 1) After introducing the text input in MASE like HierTTS, the CMOS is reduced by 0.12. We also observe that the KL loss converges faster at the word, sentence, and paragraph levels, which means that less information is compressed into these levels. 2) Removing the MSD and using only the $\bm{z}$ from the frame layer reduces CMOS by 0.26. This most significant decrease proves our hypothesis in Sec. \ref{ssec:msd}. 3) Finally, replacing our training strategy with the staged KL weighted annealing in HierTTS reduces the CMOS by 0.17. 
It is worth noting that, unlike the staged KL weighted annealing, our training strategy does not require extensive experience in deciding when to increase the KL weights of a particular level.

\section{Conclusion}
\label{sec:conclusion}
In this paper, we introduce multi-step VAE into the VITS to enhance its performance on audiobook paragraph synthesis. In addition, we propose a series of modifications to improve the multi-step VAE. Experiment results demonstrate that the paragraph speech synthesized by EP-MSTTS outperforms all the baseline models. Ablation studies confirm the effectiveness of our proposed improvements. In the future, we will further analyze the information that has been decoupled into different grammatical layers and explore to realize the smooth transition of synthesized speech between paragraphs, using coarse-grained information of the paragraph layer.

\section{Acknowledgements}
This work is supported by the National Key Research and Development Program of China(No. 2021YFC3320102) and Postdoctoral Fellowship Program of CPSF(No. GZB20230811)

\bibliographystyle{IEEEtran}
\bibliography{mybib}

\end{document}